\documentclass[12pt]{article}

\usepackage{amsmath,amsfonts,amssymb,latexsym}
\def\mco{\multicolumn}

\numberwithin{equation}{section}
\def\be{\begin{equation}}
\def\ee{\end{equation}}
\def\bq{\begin{eqnarray}}
\def\eq{\end{eqnarray}}
\def\beq{\begin{eqnarray*}}
\def\eeq{\end{eqnarray*}}

\def\a{\alpha}

\def\d{\delta}

\begin{document}
\begin{titlepage}
\begin{flushright}
CERN-PH-TH/2010-104
\end{flushright}

\vspace{0.4cm}

\begin{center}
{\LARGE Brane singularities and their avoidance in a fluid bulk}

\vspace{0.6cm}

{\large Ignatios Antoniadis$^{1,*,3}$, Spiros Cotsakis$^{2,\dagger}$, Ifigeneia Klaoudatou$^{2,\ddagger}$}\\

\vspace{0.5cm}

$^1$ {\normalsize {\em Department of Physics, CERN - Theory Division}}\\
{\normalsize {\em CH--1211 Geneva 23, Switzerland}}\\

\vspace{2mm}

$^2${\normalsize {\em Research Group of Cosmology, Geometry and
Relativity}}\\ {\normalsize {\em Department of Information and
Communication Systems Engineering}}\\
{\normalsize {\em University of the Aegean,}} 
{\normalsize {\em Karlovassi 83 200, Samos, Greece}}\\
\vspace{2mm}
{\normalsize {\em E-mails:} $^*$\texttt{ignatios.antoniadis@cern.ch},
$\dagger$\texttt{skot@aegean.gr}, $\ddagger$\texttt{iklaoud@aegean.gr}}
\end{center}

\vspace{0.1cm}

\begin{abstract}
\noindent Using the method of asymptotic splittings,  the possible
singularity structures and the corresponding asymptotic behavior of
a 3-brane in a five-dimensional bulk are classified, in the case
where the bulk field content is parametrized by an analog of perfect
fluid with an arbitrary equation of state $P=\gamma\rho$ between the
`pressure' $P$ and the `density' $\rho$. In this analogy with
homogeneous cosmologies, the time is replaced by the extra
coordinate transverse to the 3-brane, whose world-volume can have an
arbitrary constant curvature. The results depend crucially on the
constant parameter $\gamma$: (i) For $\gamma>-1/2$, the flat brane
solution suffers from a collapse singularity at finite distance,
that disappears in the curved case. (ii) For $\gamma<-1$, the
singularity cannot be avoided and it becomes of the type big rip for
a flat brane. (iii) For $-1<\gamma\le -1/2$, the surprising result
is found that while the curved brane solution is singular, the flat
brane is not, opening the possibility for a revival of the
self-tuning proposal.

\end{abstract}

\begin{center}
{\line(5,0){280}}
\end{center}
$^3${\small On leave from {\em CPHT (UMR CNRS 7644) Ecole Polytechnique, F-91128 Palaiseau}}

\end{titlepage}
\section{Introduction}

In a previous work~\cite{iklaoud_6, ack1}, we  started a general
analysis of the singularity structure and the corresponding
asymptotic behavior of a 3-brane embedded in a five-dimensional
bulk, using the powerful method of asymptotic
splittings~\cite{skot}. The main physical motivation is the
so-called fine-tuning mechanism of the cosmological constant on the
brane world-volume~\cite{nima,silver}. In Refs.~\cite{iklaoud_6,
ack1}, we considered an extended version of the simplest model
studied in the past, containing a massless bulk scalar field with a
general coupling to the brane (corresponding to an arbitrary
localized potential), and we allowed the brane world-volume to have
non-vanishing constant curvature.

We then found that the collapse singularity at a  finite distance
from the brane is present in all solutions with a flat
brane~\cite{Gubser,Forste}, but can be avoided ({\it i.e.} moved at
infinite distance) when the brane becomes curved, either positively
or negatively. The singularity in the flat case is of the big bang
type, characterized by the vanishing of the warp factor and the
divergence of its derivative, as well as of the `density' of the
scalar field.

In this paper we extend the previous analysis to the case where the
bulk matter is described by an analog of a
 perfect fluid. In fact, the case of a massless bulk scalar is a  particular case of
such a fluid, corresponding to an equation of state $P= \gamma\rho$
with $\gamma=1$. Indeed, the asymptotic behavior near the
singularity of the flat brane solution given in eq.~(3.4) of
Ref.~\cite{ack1}, is identical to eq.~(\ref{preveq}) below for
$\gamma=1$. Here, we are interested in the general dynamics of
`evolution' of such a brane-world analog to cosmology for arbitrary
$\gamma$, in order to reveal the various types of singularity that
may develop within a finite distance from the original position of
the brane, and on the other hand to determine conditions that may
lead to the avoidance of the singularities shifting them at an
infinite distance away from the brane.

In particular, we shall show that the existence of a
perfect fluid in the bulk enhances the dynamical possibilities of
brane evolution in the fluid bulk. Such possibilities stem from the
different possible behaviors of the fluid density and the
derivative of the warp factor with respect to the extra dimension.
The result depends crucially on the values of the parameter $\gamma$.
In general, we find three regions of $\gamma$ leading to qualitatively different results:
\begin{itemize}
\item In the region $\gamma>-1/2$, the situation is very similar to the case of a massless bulk scalar field. A flat brane solution has necessarily a collapse singularity at finite distance, which is moved at infinity when the brane becomes curved.
\item The situation changes drastically in the region $-1<\gamma<-1/2$. The curved brane solution becomes singular while the flat brane is regular. Thus, this region seems to avoid the main obstruction of the self-tuning proposal: any value of the brane tension is absorbed in the solution and the brane remains flat. The main question is then whether there is a field theory realization of such a fluid producing naturally an effective equation of state of this type.
\item Finally, in the region $\gamma<-1$, corresponding to  the analog of a phantom equation of state,
we show that it is possible for the brane to be ripped apart in as
much the same way as in a big  rip singularity. This happens only in
the flat case, while curved brane solutions develop a `standard'
collapse singularity. No regular solution is found in this region.
\end{itemize}
Besides the above regions, the values $\gamma=-1/2,-1$ are special:
for $\gamma=-1/2$,  we find again a regular flat brane solution with
the so-called sudden behavior~\cite{ba04}, as well as a non-singular
curved brane, while for $\gamma=-1$ there is only singular curved
solution.

As mentioned above, it would be very interesting to  understand
whether there are field theory representations reproducing the
`exotic' regions of $\gamma\le -1/2$. Obviously the analogy of the
perfect fluid concerning the positivity energy conditions does not
seem to apply in this case where time is replaced by an additional
space coordinate. However, some restrictions may be applied from
usual field theory axioms. Also, the formation of singularities
discussed here is better understood in a dynamical rather than the
usual geometric sense met in general relativity. In the latter case,
cosmological singularities are forming together with conjugate (or focal) points
in spacetime, and for this it is necessary that there exists at
least one timelike dimension (and any number of spacelike ones,
greater than two). The timelike dimension forces then the geodesics
to focus along it rather than along any of the spacelike dimensions.
In the problem discussed in this paper, the timelike dimension is on
the brane while the singularities are forming along spacelike
dimensions in the bulk. As we show below these singularities are
real in the dynamical sense that some component of the solution
vector $(a, a', \rho)$ diverges there. Therefore we abandon the
usual interpretation according to which the universe comes to an end
in a finite time possibly  through geodesic refocusing, and instead
we study how dynamical effects guide our brane systems to extreme
behaviors.

The structure of this paper is the following. In Section 2, we first
choose appropriate variables and rewrite the basic field equations
of the problem in the form of a dynamical system; secondly, we
introduce some convenient terminology for the different types of
singularity to be met later in our analysis; thirdly, we single out
the possible dominant balances, organizing centers of all the
evolutionary behaviors that fully characterize our problem. In
Sections 3 and 4, we study carefully the asymptotics around collapse
singularities of two types, that we call I and II, respectively. In
Section 5, we explore the dynamics as the brane approaches a big rip
singularity, while in Section 6 we look at a milder singularity that
resembles in many ways the so-called sudden (non-singular) behavior
introduced in Ref. \cite{ba04}. In Section 7, we analyze the
possibility of avoiding finite-distance singularities leading to the
existence of regular brane evolution in the bulk and finally, in
Section 8 we conclude and refer to future work.
\section{Dynamics in a perfect fluid bulk}
In this Section we rewrite the brane model living in a bulk filled
with a perfect fluid as a dynamical system in three basic variables
and completely identify the principal modes of approach to its
singularities, that is we find all the dominant balances of the
system. We consider a three-brane embedded in a five-dimensional
bulk space filled with a perfect fluid with equation of state
$P=\gamma \rho$, where the pressure $P$ and the density $\rho$ are
functions only of the fifth dimension, denoted by $Y$. We assume a
bulk metric of the form \be \label{warpmetric}
g_{5}=a^{2}(Y)g_{4}+dY^{2}, \ee where $g_{4}$ is the
four-dimensional flat, de Sitter or anti de Sitter metric, i.e., \be
\label{branemetrics} g_{4}=-dt^{2}+f^{2}_{\kappa}g_{3}, \ee where
\be g_{3}=dr^{2}+h^{2}_{\kappa}g_{2} \ee and \be
g_{2}=d\theta^{2}+\sin^{2}\theta d\varphi^{2}. \ee Here
$f_{\kappa}=r,\sin r,\sinh r, $ and $ h_{\kappa}=1,\cosh (H
t)/H,\cos (H t)/H $ ($H^{-1}$ is the de Sitter curvature radius). We
also assume an energy-momentum tensor of the form
$T_{AB}=(\rho+P)u_{A}u_{B}-Pg_{AB}$, where $A,B=1,2,3,4,5$ and
$u_{A}=(0,0,0,0,1)$, with the 5th coordinate corresponding to $Y$.

The five-dimensional Einstein equations,
\be
G_{AB}=\kappa^{2}_{5}T_{AB},
\ee
with $\kappa^{2}_{5}=M_{5}^{-3}$ and $M_{5}$ the five dimensional Planck
mass, can be written in the following form:
\bq
\label{syst2i}
\frac{a''}{a}&=&-\kappa^{2}_{5}\frac{(1+2\gamma)}{6}\rho, \\
\label{syst2iii} \frac{a'^{2}}{a^{2}}&=&\frac{\kappa^{2}_{5}}{6} A
\rho+\frac{k H^{2}}{a^{2}}, \eq where $k=0,\pm 1$, and the prime
($'$) denotes differentiation with respect to $Y$. The equation of
conservation, \be \nabla_{B}T^{AB}=0, \ee becomes, \be
\label{syst2ii} \rho'+4(1+\gamma)H\rho=0. \ee Introducing the new
variables \be x=a, \quad y=a', \quad z=\rho, \ee Eqs. (\ref{syst2i})
and (\ref{syst2ii}) take the form \bq \label{syst2a}
x'&=&y, \\
y'&=&-2A\frac{(1+2\gamma)}{3}z x, \\
\label{syst2c} z'&=&-4(1+\gamma)\frac{y}{x}z, \eq while eq.~(\ref{syst2iii}) reads \be \label{constraint3}
\frac{y^{2}}{x^{2}}=\frac{2}{3} A z+\frac{k H^{2}}{x^{2}}, \quad
A=\kappa_{5}^{2}/4. \ee Since this last equation does not contain
derivatives with respect to $Y$, it is a velocity independent
constraint equation for the system (\ref{syst2a})-(\ref{syst2c}).
Before we proceed with the analysis of the above system, we
introduce the following terminology for the possible singularities
to occur at a finite-distance from the brane. Specifically, we call
a state where:
\begin{enumerate}
\item[i)]$  $ $a\rightarrow 0$, $a'\rightarrow \infty$ and
$\rho\rightarrow \infty$: a singularity of collapse type I.
\item[ii)]$  $ $a\rightarrow 0$, $a'\rightarrow a'_{s}$ and
$\rho\rightarrow 0$: a singularity of collapse type IIa,\\
\indent$a\rightarrow 0$, $a'\rightarrow a'_{s}$ and
$\rho\rightarrow \rho_{s}$: a singularity of collapse type IIb,\\
\indent$a\rightarrow 0$, $a'\rightarrow a'_{s}$ and
$\rho\rightarrow \infty$: a singularity of collapse type IIc,\\
where $a'_{s}$, $\rho_{s}$ are non-vanishing finite constants.
\item[iii)]$  $ $a\rightarrow \infty$, $a'\rightarrow -\infty$ and
$\rho\rightarrow\infty$: a big rip singularity.
\end{enumerate}
We note that with this terminology the finite-distance singularity
studied in \cite{ack1} is a singularity of collapse type I.

The next step is to apply the method of asymptotic splittings in an
effort to find all possible asymptotic behaviors of the dynamical
system (\ref{syst2a})-(\ref{syst2c}) with the constraint
(\ref{constraint3}), by building series expansions of the solutions
around the presumed position of the singularity at $Y_{s}$.

We note that the system (\ref{syst2a})-(\ref{syst2c}) is a weight homogeneous
system determined by the vector field
\be
\label{vectorfield}
\mathbf{f}=\left(y,-2A\frac{(1+2\gamma)}{3}z x,-4(1+\gamma)
\frac{y}{x}z\right)^{\intercal}.
\ee

In order to compute all possible dominant balances that describe the
principal asymptotics of the system we look for pairs of the form,
\be
\mathcal{B}=\{\mathbf{a},\mathbf{p}\}, \quad \textrm{where} \quad
\mathbf{a}=(\alpha,\beta,\delta), \quad \mathbf{p}=(p,q,r),
\ee
with
\be
(p,q,r)\in\mathbb{Q}^{3} \quad \textrm{and} \quad
(\alpha,\beta,\delta)\in \mathbb{C}^{3}\smallsetminus\{\mathbf{0}\},
\ee
by setting $(x,y,z)=(\alpha\Upsilon^{p},\beta \Upsilon^{q},\delta
\Upsilon ^{s})$ in the system (\ref{syst2a})-(\ref{syst2c}), where $\Upsilon=Y-Y_s$ is the distance from the singularity. We find after some
calculation the following list of all possible balances for our basic system
(\ref{syst2a})-(\ref{constraint3}):
\bq
_{\gamma}\mathcal{B}_{1}&=&
\left\{\left(\alpha,\alpha p,\frac{3}{2A}p^{2}\right),(p,p-1,-2)\right\},
p=\frac{1}{2(\gamma+1)}, \, \gamma \neq -1/2,-1,\label{preveq}\\
_{\gamma}\mathcal{B}_{2}&=&\{(\alpha,\alpha,0),(1,0,-2)\},\quad \gamma \neq -1/2,\\
_{-1/2}\mathcal{B}_{3}&=&\{(\alpha,\alpha,0),(1,0,r)\},
\\
_{-1/2}\mathcal{B}_{4}
&=&\{(\alpha,\alpha,\delta),(1,0,-2)\},\\
_{-1/2}\mathcal{B}_{5}&=&\{(\a,0,0), (0,-1,r)\},\label{Bfive} \eq where
$_{-1/2}\mathcal{B}_{i}\equiv_{\gamma=-1/2}\mathcal{B}_{i}$.
Notice that, as already mentioned in the introduction, the first balance
$_{\gamma}\mathcal{B}_{1}$ for $\gamma=1$ coincides with the one found in \cite{ack1}
in eq.~(3.4), where the fluid was replaced by a massless bulk scalar field
with an arbitrary coupling to the brane.

The above balances are \emph{exact} solutions of the system and they must
therefore also satisfy the constraint equation (\ref{constraint3}).
This fact alters the presumed generality of the solution represented
by each one of the balances above and determines uniquely the type
of spatial geometry that we must consider: The balances
$_{\gamma}\mathcal{B}_{1}$ and $_{-1/2}\mathcal{B}_{5}$ are found
when we set $k=0$, and describe a (potentially general) solution
corresponding to a flat brane, while the balances
$_{\gamma}\mathcal{B}_{2}$ and $_{-1/2}\mathcal{B}_{3}$ were found
when $k\neq 0$ and describe particular solutions of curved branes
(since we already have to sacrifice the arbitrary constant $\a$ by
imposing $\a^{2}=kH^{2}$). For the balance
$_{-1/2}\mathcal{B}_{4}$, on the other hand, $k$ is not specified and
hence it describes a particular solution for a curved or flat brane
(particularly since we have to set $\d=(3/(2A))(1-kH^{2}/\a^{2})$ to
satisfy eq.~(\ref{constraint3})).

Each one of these balances are analyzed in detail in the following sections
according to the nature of asymptotic behaviors they imply.
\section{Collapse type I singularity}
We shall focus in this Section exclusively on the balance
$_{\gamma}\mathcal{B}_{1}$ and show that for certain ranges of $\gamma$ it
gives the generic asymptotic behavior of a flat brane to a singularity of
collapse type I. Our analysis implies that such behavior can only result from a
$_{\gamma}\mathcal{B}_{1}$ type of balance.

Our purpose is to construct asymptotic expansions of solutions to
the dynamical system (\ref{syst2a})-(\ref{constraint3}) in the form
of a series solution defined by \be \label{Puiseux}
\mathbf{x}=\Upsilon^{\mathbf{p}}(\mathbf{a}+
\Sigma_{j=1}^{\infty}\mathbf{c}_{j}\Upsilon^{j/s}), \ee where
$\mathbf{x}=(x,y,z)$, $\mathbf{c}_{j}=(c_{j1},c_{j2},c_{j3})$, and
$s$ is in this case the least common multiple of the denominators of
the positive $\mathcal{K}$-exponents (cf. \cite{skot},
\cite{goriely}). As a first step we calculate the Kowalevskaya
matrix,
$\mathcal{K}=D\mathbf{f}(\mathbf{a})-\textrm{diag}(\mathbf{p})$,
where $D\mathbf{f}(\mathbf{a})$ is the Jacobian matrix of
$\mathbf{f}$ \footnote{$\mathbf{f}$ is the vector field resulting
from the dynamical system (\ref{syst2a})-(\ref{constraint3}) and
$\{\mathbf{a},\mathbf{p}\}$ is the balance
$_{\gamma}\mathcal{B}_{1}$.}: We have, \be D\mathbf{f}(x,y,z)=\left(
                     \begin{array}{ccc}
                       0 & 1             & 0 \\ \\
               -\dfrac{2}{3}(1+2\gamma)A z & 0             & -\dfrac{2}{3}(1+2\gamma)A x  \\ \\
       4(1+\gamma)\dfrac{y z}{x^{2}} & -4(1+\gamma)\dfrac{z}{x} & -4(1+\gamma)\dfrac{y}{x} \\
                     \end{array}
                   \right),
\ee
to be evaluated on $\mathbf{a}$. The balance $_{\gamma}\mathcal{B}_{1}$ has
$\mathbf{a}=(\alpha,\alpha p,3p^{2}/2A)$, and $\mathbf{p}=(p,p-1,-2)$, with
$p=1/(2(\gamma+1))$. Thus the Kowalevskaya matrix ($\mathcal{K}$-matrix in short)
for this balance, 
is
\beq
\quad\quad\quad_{\gamma}\mathcal{K}_{1}&=&D\mathbf{f}\left(\a,\a p,\frac{3}{2A}p^{2}\right)-\textrm{diag}(p,p-1,-2)\\
&=&
D\mathbf{f}\left(a,\frac{a}{2(1+\gamma)},\frac{3}{8A(1+\gamma)^{2}}\right)
-\textrm{diag}\left(\frac{1}{2(1+\gamma)},-\frac{1+2\gamma}{2(1+\gamma)},-2\right)
\eeq
\be =\left(
                     \begin{array}{ccc}
                       -\dfrac{1}{2(1+\gamma)} & 1             & 0 \\ \\
               -\dfrac{1+2\gamma}{4(1+\gamma)^{2}} & \dfrac{1+2\gamma}{2(1+\gamma)}           & -\dfrac{2}{3}(1+2\gamma)A \a  \\ \\
       \dfrac{3}{4(1+\gamma)^{2}A \a} & -\dfrac{3}{2(1+\gamma)A \a} & 0 \\
                     \end{array}
                   \right).
\ee
\\

Let us now calculate what the $\mathcal{K}$-exponents for this balance actually
are. Recall that these exponents
are the eigenvalues of the matrix $_{\gamma}\mathcal{K}_{1}$ and
constitute its spectrum, $spec(_{\gamma}\mathcal{K}_{1})$.
The arbitrary constants of any (particular or general) solution first appear in
those terms whose coefficients $\mathbf{c}_{k}$ have indices $k=\varrho s$, where $\varrho$
is a non-negative $\mathcal{K}$-exponent. The number of non-negative $\mathcal{K}$-exponents equals therefore
the number of arbitrary constants that appear in the series expansions
of (\ref{Puiseux}). There is always the $-1$ exponent that corresponds to an
arbitrary constant that is the position of the singularity at $Y_{s}$. The
balance $_{\gamma}\mathcal{B}_{1}$ corresponds thus to a general solution in our
case if and only if it possesses two non-negative $\mathcal{K}$-exponents
(the third arbitrary constant is the position of the singularity, $Y_{s}$).
Here we find
\be
\textrm{spec}(_{\gamma}\mathcal{K}_{1})=
\left\{-1,0,\frac{1+2\gamma}{1+\gamma}\right\}.
\ee
The last eigenvalue is a function of the $\gamma$ parameter and it is
positive when either $\gamma<-1$, or $\gamma>-1/2$. We consider here the case
$\gamma>-1/2$ since, as it will soon follow, this
range of $\gamma$ is adequate for the occurrence of a collapse type I singularity.
The case of $\gamma<-1$ leads to a big rip singularity and will be
examined in Section 5.

Let us assume $\gamma=-1/4$ for concreteness. Then
\bq
_{-1/4}\mathcal{B}_{1}&=&\{(\a,-2\a/3,2/(3A)),(2/3,-1/3,-2)\},\\
\textrm{spec}(_{-1/4}\mathcal{K}_{1})&=&\{-1,0,2/3\}.
\eq
Substituting in the system (\ref{syst2a})-(\ref{syst2c}) the
particular value $\gamma=-1/4$ and the forms
\be
x=\Sigma_{j=0}^{\infty}c_{j1}\Upsilon^{j/3+2/3}, \quad
y=\Sigma_{j=0}^{\infty}c_{j2}\Upsilon^{j/3-1/3}, \quad
z=\Sigma_{j=0}^{\infty}c_{j3}\Upsilon^{j/3-2},
\ee
we arrive at the following asymptotic expansions:
\bq
\label{g_B1x}
x&=& \a\Upsilon^{2/3}-\frac{A\a}{2}c_{2\,3}\Upsilon^{4/3}+\cdots,\\
y&=& \frac{2}{3}\a\Upsilon^{-1/3}-\frac{2}{3}A \a c_{2\,3}\Upsilon^{1/3}+\cdots,\\
\label{g_B1w} z&=&
\frac{2}{3A}\Upsilon^{-2}+c_{2\,3}\Upsilon^{-4/3}+\cdots. \eq For
this to be a valid solution we need to check whether for each $j$
satisfying $j/3=\varrho$ with $\varrho$ a positive eigenvalue, the
corresponding eigenvector $v$ of the $_{-1/4}\mathcal{K}_{1}$ matrix
is such that the compatibility conditions hold, namely, we must have
\be v^{\top}\cdot P_{j}=0, \ee where $P_{j}$ are polynomials in
$\mathbf{c}_{i},\ldots, \mathbf{c}_{j-1}$ given by \be
_{-1/4}\mathcal{K}_{1}\mathbf{c}_{j}-(j/3)\mathbf{c}_{j}=P_{j}. \ee
Here the corresponding relation $j/3=2/3$, is valid only for $j=2$
and the compatibility condition
indeed holds since,
\be (_{-1/4}\mathcal{K}_{1}-(2/3)\mathcal{I}_{3})\mathbf{c}_{2}= \left(
  \begin{array}{ccc}
    -\dfrac{4}{3} & 1 & 0 \\ \\
    -\dfrac{2}{9} & -\dfrac{1}{3} & -\dfrac{A\a}{3} \\ \\
    \dfrac{4}{3A\a} & -\dfrac{2}{A\a} & -\dfrac{2}{3} \\
  \end{array}
\right)c_{2\,3}\left(
                 \begin{array}{c}
                   -\dfrac{A\a}{2} \\ \\
                   -\dfrac{2A\a}{3}  \\ \\
                   1 \\
                 \end{array}
               \right)=\left(
                         \begin{array}{c}
                           0 \\ \\
                           0 \\ \\
                           0 \\
                         \end{array}
                       \right).
\ee
Eqs. (\ref{g_B1x})-(\ref{g_B1w}) then imply that as
$\Upsilon\rightarrow 0$, \be a\rightarrow 0, \quad a'\rightarrow
\infty, \quad \rho\rightarrow \infty.
\ee
This asymptotic behavior corresponds to a general solution of a flat
brane that is valid around a collapse I singularity.
We thus 
regain a behavior similar to the one met in \cite{ack1} for the case of a flat
brane in a scalar field bulk.
\section{Collapse type II singularities}
In this Section, we show that for a curved brane ($k=\pm 1$) the
long-term (distance) behavior of all solutions which depend on the
asymptotics near finite-distance singularities turn out to be of a
very different nature. In particular, we shall show that the
balances $_{\gamma}\mathcal{B}_{2}$ for $\gamma<-1/2$,
$_{-1/2}\mathcal{B}_{3}$ for $r<-2$ and $_{-1/2}\mathcal{B}_{4}$ (as
we have already mentioned $_{\gamma}\mathcal{B}_{2}$ and
$_{-1/2}\mathcal{B}_{3}$ correspond to a curved brane whereas
$_{-1/2}\mathcal{B}_{4}$ corresponds to a flat or curved brane),
imply the existence of a collapse type IIa, b or c singularity. This
is in sharp contrast to the asymptotic behavior found for a curved
brane in the presence of a bulk scalar field (cf. \cite{ack1}), wherein there
are no finite-distance singularities.

For the balance $_{\gamma}\mathcal{B}_{2}$ we find that
\be
_{\gamma}\mathcal{K}_{2}=D\mathbf{f}\left(\a,\a,0\right)
-\textrm{diag}\left(1,0,-2\right)= \left(
  \begin{array}{ccc}
    -1 & 1 & 0 \\
    0 & 0 & -\dfrac{2}{3}A\a (1+2\gamma)\\
    0 & 0 & -2(1+2\gamma) \\
  \end{array}
\right),
\ee
and hence,
\be
\textrm{spec}(_{\gamma}\mathcal{K}_{2})=\{-1,0,-2(1+2\gamma)\}.
\ee
We note that the third arbitrary constant appears at the value
$j=-2(1+2\gamma)$, $\gamma<-1/2$. After substituting the forms,
\be
x=\Sigma_{j=0}^{\infty}c_{j1}\Upsilon^{j+1},\quad
y=\Sigma_{j=0}^{\infty}c_{j2}\Upsilon^{j},\quad
z=\Sigma_{j=0}^{\infty}c_{j3}\Upsilon^{j-2},
\ee
in the system (\ref{syst2a})-(\ref{syst2c}), to proceed
we may try giving different values to $\gamma$: Inserting the value $\gamma
=-3/4$ in the system for concreteness we meet a third arbitrary constant at $j=1$
($\textrm{spec}(_{-3/4}\mathcal{K}_{2})=\{-1,0,1\}$).
We then arrive at the following asymptotic forms of the solution:
\bq
\label{-3/4_B2x}
x&=& \a\Upsilon+\frac{A\a}{6}c_{1\,3}\Upsilon^{2}+\cdots,\\
y&=& \a+\frac{A\a}{3} c_{1\,3}\Upsilon+\cdots,\\
\label{-3/4_B2w} z&=&  c_{1\,3}\Upsilon^{-1}+\cdots, \eq where
$c_{1\,3}\neq 0$ \footnote{If we do not set from the beginning
$\gamma=-3/4$ but instead we let $\gamma$ be arbitrary, then in the
last step of the calculations at the $j=1$ level we find that either
$c_{1\,3}=0$ or $\gamma=-3/4$.}. We need to check the validity of
the compatibility condition for $j=1$. But this is trivially
satisfied since, \be
(_{-3/4}\mathcal{K}_{2}-\mathcal{I}_{3})\mathbf{c}_{1}= \left(
  \begin{array}{ccc}
    -2 & 1  & 0 \\
     0 & -1 & A\a/3 \\
     0 & 0  & 0 \\
  \end{array}
\right)c_{1\,3}\left(
                 \begin{array}{c}
                   A\a/6 \\
                   A\a/3  \\
                   1 \\
                 \end{array}
               \right)=\left(
                         \begin{array}{c}
                           0 \\
                           0 \\
                           0 \\
                         \end{array}
                       \right).
\ee

The series expansions in eqs.~(\ref{-3/4_B2x})-(\ref{-3/4_B2w}) are
therefore valid and we conclude that as $\Upsilon\rightarrow
0$,
\be
\label{-3/4B2} a\rightarrow 0, \quad a'\rightarrow\a, \quad
\rho\rightarrow \infty, \quad \alpha\neq 0.
\ee
This is a collapse type IIc singularity. It will follow from the analysis
below that the behavior of $\rho$ depends on our choice of $\gamma$
(thus giving rise to three possible subcases of a type II singularity).
Indeed, choosing for instance $\gamma=-1$ ($\textrm{spec}(_{-1}\mathcal{K}_{2})=\{-1,0,2\}$),
we find that the solution is given by the forms,
\bq
\label{-1_B2x}
x&=& \a\Upsilon+\frac{A\a}{9}c_{2\,3}\Upsilon^{3}+\cdots,\\
y&=& \a+\frac{A\a}{3} c_{2\,3}\Upsilon^{2}+\cdots,\\
\label{-1_B2w} z&=&  c_{2\,3}+\cdots,
\eq
where $c_{2\,3}\neq 0$ \footnote{Had we let $\gamma$ be arbitrary we would have
found that in the step $j=2$ of the procedure either $c_{2\,3}=0$ or
$\gamma=-1$.}.
Note that the compatibility condition is satisfied here as well since,
\be
(_{-1}\mathcal{K}_{2}-2\mathcal{I}_{3})\mathbf{c}_{2}= \left(
  \begin{array}{ccc}
    -3 & 1  & 0 \\
     0 & -2 & 2A\a/3 \\
     0 & 0  & 0 \\
  \end{array}
\right)c_{2\,3}\left(
                 \begin{array}{c}
                   A\a/9 \\
                   A\a/3 \\
                   1 \\
                 \end{array}
               \right)=\left(
                         \begin{array}{c}
                           0 \\
                           0 \\
                           0 \\
                         \end{array}
                       \right).
\ee
We see that as $\Upsilon\rightarrow 0$, \be a\rightarrow 0,
\quad a'\rightarrow\a, \quad \rho\rightarrow c_{2\,3}, \quad \a\neq
0.
\ee
This is a collapse type IIb singularity in our terminology and is clearly
different from (\ref{-3/4B2}).

A yet different behavior is met if we choose for instance $\gamma=-5/4$. The
$\mathcal{K}$-exponents are given by
$\textrm{spec}(_{-5/4}\mathcal{K}_{2})=\{-1,0,3\}$, and the series
expansions become,
\bq
\label{-5/4_B2x}
x&=& \a\Upsilon+\frac{A\a}{12}c_{3\,3}\Upsilon^{4}+\cdots,\\
y&=& \a+\frac{A\a}{3} c_{3\,3}\Upsilon^{3}+\cdots,\\
\label{-5/4_B2w} z&=&  c_{3\,3}\Upsilon+\cdots, .
\eq
where $c_{3\,3}\neq 0$ \footnote{Here again if had let $\gamma$ be arbitrary we
would have found that in the step $j=3$ of the procedure either $c_{3\,3}=0$ or
$\gamma=-5/4$.}.
These expansions are valid locally around the singularity since the
compatibility condition holds true because,
\be
(_{-5/4}\mathcal{K}_{2}-3\mathcal{I}_{3})\mathbf{c}_{1}= \left(
  \begin{array}{ccc}
    -4 & 1  & 0 \\
     0 & -3 & A\a \\
     0 & 0  & 0 \\
  \end{array}
\right)c_{3\,2}\left(
                 \begin{array}{c}
                   A\a/12 \\
                   A\a/3  \\
                   1 \\
                 \end{array}
               \right)=\left(
                         \begin{array}{c}
                           0 \\
                           0 \\
                           0 \\
                         \end{array}
                       \right).
\ee
For $\Upsilon\rightarrow 0$, we have that
\be
a\rightarrow 0,
\quad a'\rightarrow\a, \quad \rho\rightarrow 0, \quad \a\neq 0,
\ee
which means that this is a collapse type IIa singularity.
This balance therefore leads to the asymptotic behavior of a
particular solution describing a curved brane approaching a collapse type II
singularity, i.e., $a\rightarrow 0$ and
$a'\rightarrow \alpha$.
The behavior of the density of the perfect fluid varies
dramatically: we can have an infinite density, a constant density, or even
no flow of ``energy" at all as we approach the finite-distance singularity into
the extra dimension at $Y_{s}$, depending on the values of
the $\gamma$ parameter. 

We now turn to an analysis of the balances $_{-1/2}\mathcal{B}_{3}$, for $r<-2$, and
$_{-1/2}\mathcal{B}_{4}$. The $\mathcal{K}$-matrix for $_{-1/2}\mathcal{B}_{3}$ is
\be
_{-1/2}\mathcal{K}_{3}=D\mathbf{f}\left(\a,\a,0\right)
-\textrm{diag}(1,0,r)= \left(
  \begin{array}{ccc}
    -1 & 1 & 0 \\
    0 & 0 & 0\\
    0 & 0 & -2-r \\
  \end{array}
\right),
\ee
and hence,
\be
\textrm{spec}(_{-1/2}\mathcal{K}_{3})=\{-1,0,-2-r\}.
\ee
Taking $-2-r>0$, we have two non-negative $\mathcal{K}$-exponents.
(The case $-2-r<0$ is considered later, in Section 8, since it is quite different, it
does not imply the existence of a finite-distance singularity.) For $r=-3$ as an example, we
substitute the forms
\be
x=\Sigma_{j=0}^{\infty}c_{j1}\Upsilon^{j+1},\quad
y=\Sigma_{j=0}^{\infty}c_{j2}\Upsilon^{j},\quad
z=\Sigma_{j=0}^{\infty}c_{j3}\Upsilon^{j-3},
\ee
and arrive at the expansions
\bq
\label{-1/2_B_3x}
x&=& \a\Upsilon+\cdots,\\
y&=& \a+\cdots,\\
\label{-1/2_B_3w} z&=&  c_{1\,3}\Upsilon^{-2}+\cdots.
\eq
The compatibility condition is satisfied because
\be
(_{-1/2}\mathcal{K}_{3}-\mathcal{I}_{3})\mathbf{c}_{1}= \left(
  \begin{array}{ccc}
    -2 & 1  & 0 \\
     0 & -1 & 0 \\
     0 & 0  & 0 \\
  \end{array}
\right)c_{1\,3}\left(
                 \begin{array}{c}
                   0 \\
                   0  \\
                   1 \\
                 \end{array}
               \right)=\left(
                         \begin{array}{c}
                           0 \\
                           0 \\
                           0 \\
                         \end{array}
                       \right),
\ee
and so the expansions (\ref{-1/2_B_3x})-(\ref{-1/2_B_3w}) are
valid ones in the vicinity of the singularity. 
The general behavior of the solution is then characterized by the
asymptotic forms \be a\rightarrow 0, \quad a'\rightarrow \a,
\quad\rho \rightarrow\infty, \quad \a\neq 0. \ee The balance
$_{-1/2}\mathcal{B}_{3}$ for $r<-2$ implies therefore the existence
of a collapse type IIc singularity during the dynamical evolution of
the curved brane living (and moving) in this specific perfect fluid
bulk.

The balance $_{-1/2}\mathcal{B}_{4}$ on the other hand is one with
\be
_{-1/2}\mathcal{K}_{4}=D\mathbf{f}\left(\a,\a,\d\right)
-\textrm{diag}(1,0,-2)= \left(
  \begin{array}{ccc}
    -1 & 1 & 0 \\
    0 & 0 & 0\\
    \dfrac{2\d}{\a} & -\dfrac{2\d}{\a} & 0 \\
  \end{array}
\right),
\ee
and
\be
\textrm{spec}(_{-1/2}\mathcal{K}_{4})=\{-1,0,0\}.
\ee
We note that the double multiplicity of the zero eigenvalue reflects the fact
that there were already two arbitrary constants, $\a$ and $\d$ in
this balance (recall though that $\d$ had to be sacrificed in order for this
balance to satisfy the constraint (\ref{constraint3})). We can thus write
\bq
\label{-1/2_B4x}
x&=& \a\Upsilon+\cdots,\\
y&=& \a+\cdots,\\
\label{-1/2_B4w} z&=&  \d\Upsilon^{-2}+\cdots, \eq so that as
$\Upsilon\rightarrow 0$, a collapse type IIc singularity develops,
i.e., \be a\rightarrow 0,\quad a'\rightarrow \a,
\quad\rho\rightarrow\infty, \quad \alpha\neq 0. \ee
\section{Big rip singularities}
In this Section we return to the balance $_{\gamma}\mathcal{B}_{1}$
but focus on different $\gamma$ values. In particular, we show that when
$\gamma<-1$, a flat brane develops a big rip singularity in a finite distance.
This new asymptotic behavior implied by the balance $_{\gamma}\mathcal{B}_{1}$
(when $\gamma<-1$) is equally general to the one found in Section 3.

For purposes of illustration, let us take $\gamma=-2$. Then the balance
$_{-2}\mathcal{B}_{1}$ and the $_{-2}\mathcal{K}_{1}$-exponents read, respectively,
\bq
_{-2}\mathcal{B}_{1}&=&\{(\a,-\a/2,3/(8A)),(-1/2,-3/2,-2)\},\\
\textrm{spec}(_{-2}\mathcal{K}_{1})&=&\{-1,0,3\}.
\eq
Substituting the value $\gamma=-2$ in our basic system given by eqs.~(\ref{syst2a})-(\ref{syst2c}), and also the forms
\be
x=\Sigma_{j=0}^{\infty}c_{j1}\Upsilon^{j-1/2},\quad
y=\Sigma_{j=0}^{\infty}c_{j2}\Upsilon^{j-3/2},\quad
z=\Sigma_{j=0}^{\infty}c_{j3}\Upsilon^{j-2},
\ee
we expect to meet the third arbitrary constant at $j=3$. Indeed we find:
\bq
\label{ph_B1x}
x&=& \a\Upsilon^{-1/2}+\frac{2}{3}A\a c_{3\,3}\Upsilon^{5/2}+\cdots,\\
y&=& -\frac{\a}{2}\Upsilon^{-3/2}+\frac{5}{3}A \a c_{3\,3}\Upsilon^{3/2}+\cdots,\\
\label{ph_B1w} z&=&
\frac{3}{8A}\Upsilon^{-2}+c_{3\,3}\Upsilon+\cdots, \quad c_{3\,3}\neq 0.
\eq
The compatibility condition is trivially satisfied for $j=3$, since the product
$(_{-2}\mathcal{K}_{1}-3\mathcal{I}_{3})\mathbf{c}_{3}$ is identically zero:
\be
(_{-2}\mathcal{K}_{1}-3\mathcal{I}_{3})\mathbf{c}_{3}= \left(
  \begin{array}{ccc}
    -\dfrac{5}{2} & 1 & 0 \\ \\
    \dfrac{3}{4} & -\dfrac{3}{2} & 2A\a \\ \\
    \dfrac{3}{4A\a} & \dfrac{3}{2A\a} & -3 \\
  \end{array}
\right)c_{3\,3}\left(
                 \begin{array}{c}
                   \dfrac{2}{3}A\a \\ \\
                   \dfrac{5}{3}A\a  \\ \\
                   1 \\
                 \end{array}
               \right)=\left(
                         \begin{array}{c}
                           0 \\
                           0 \\
                           0 \\
                         \end{array}
                       \right).
\ee
The series expansions given by eqs.~(\ref{ph_B1x})-(\ref{ph_B1w})
are therefore valid asymptotically for $\Upsilon\rightarrow 0$ so
that we end up with the asymptotic forms
\be
a\rightarrow \infty,\quad a'\rightarrow -\infty,
\quad\rho\rightarrow \infty.
\ee
We therefore conclude that the balance $_{\gamma}\mathcal{B}_{1}$ 
leads to a general solution in which a flat brane develops a big rip
singularity after `traveling' for a finite distance when the bulk
perfect fluid satisfies a phantom-like equation of state, i.e.,
$\gamma<-1$. Note that using the analogy between the warp factor of
our braneworld and the scale factor of an expanding universe, we can
say that this singularity bares many similarities to the one studied in
Refs. \cite{cald99}, \cite{cald03}, \cite{go03}, since it is also
characterized by all quantities $a$, $a'$, $\rho$, and consequently
$P$, becoming asymptotically divergent. Thus, the results in this Section
indicate that a flat brane traveling in a $\gamma<-1$ fluid bulk
develops a big rip singularity. This implements the
behavior found in Section 3 of the present paper, wherein the same
brane moving in a $\gamma>-1/2$ fluid bulk `disappears' in a
big bang-type singularity.
\section{Sudden behavior}
As our penultimate mode of approach to the finite-distance
singularity, we examine here the balance
$_{-1/2}\mathcal{B}_{5}=\{(\a,0,0), (0,-1,r)\}$. This balance has
\be _{-1/2}\mathcal{K}_{5}=D\mathbf{f}\left(\a,0,0\right)
-\textrm{diag}(0,-1,r)= \left(
  \begin{array}{ccc}
    0 & 1 & 0 \\
    0 & 1 & 0\\
    0 & 0 & -r \\
  \end{array}
\right), \ee and \be
\textrm{spec}(_{-1/2}\mathcal{K}_{5})=\{1,0,-r\}, \ee so we shall
have to set $r=1$ in order to have the necessary $-1$ eigenvalue
corresponding to the arbitrary position of the ``singularity",
$Y_{s}$. After substitution of the forms \be
x=\Sigma_{j=0}^{\infty}c_{j1}\Upsilon^{j},\quad
y=\Sigma_{j=0}^{\infty}c_{j2}\Upsilon^{j-1},\quad
z=\Sigma_{j=0}^{\infty}c_{j3}\Upsilon^{j+1}, \ee we find that the
solution reads \bq \label{-1/2_B4xn}
x&=& \a+c_{1\,1}\Upsilon+\cdots,\\
y&=& c_{1\,1}+\cdots,\\
\label{-1/2_B4wn}
z&=&  0+\cdots.
\eq
The compatibility condition is satisfied since
\be
(_{-1/2}\mathcal{K}_{5}-\mathcal{I}_{3})\mathbf{c}_{1}=
\left(
  \begin{array}{ccc}
    -1 & 1  & 0 \\
     0 & 0  & 0 \\
     0 & 0  & 0 \\
  \end{array}
\right)c_{1\,1}\left(
                 \begin{array}{c}
                   1 \\
                   1 \\
                   0 \\
                 \end{array}
               \right)=\left(
                         \begin{array}{c}
                           0 \\
                           0 \\
                           0 \\
                         \end{array}
                       \right),
\ee
and we see that as $\Upsilon\rightarrow 0$,
\be
x\rightarrow \a, \quad y\rightarrow c_{1\,1}, \quad z\rightarrow 0,
\quad \alpha\neq 0.
\ee
This clearly indicates that the brane experiences the so-called sudden behavior
(cf. \cite{ba04}).
\section{Behavior at infinity}
A totally different picture than what we have already encountered up
to now in our analysis of brane singularities in a fluid bulk, is
attained using the balance $_{\gamma}\mathcal{B}_{1}$ with
$-1<\gamma<-1/2$, or the balance $_{\gamma}\mathcal{B}_{2}$ with
$\gamma >-1/2$, or the balance $_{-1/2}\mathcal{B}_{3}$ with $r>-2$.
We show in this section that these three balances and only these
offer the possibility of avoiding the finite-distance singularities
met before and may describe the behavior of our model at infinity.

We begin with the balance $_{\gamma}\mathcal{B}_{1}$ when
$-1<\gamma<-1/2$. Choosing for instance $\gamma =-4/5$, we find
$\textrm{spec}(_{-4/5}\mathcal{K}_{1})=\{-1,0,-3\}$ and
hence we may expand $(x,y,z)$ in descending powers in order to meet the
arbitrary constant appearing at $j=-1$ and $j=-3$, i.e.,
\be
x=\Sigma_{j=0}^{-\infty}c_{j1}\Upsilon^{j+5/2},\quad
y=\Sigma_{j=0}^{-\infty}c_{j2}\Upsilon^{j+3/2},\quad
z=\Sigma_{j=0}^{-\infty}c_{j3}\Upsilon^{j-2}.
\ee
We find:
\bq
\label{-4/5_B1x}
x&=& \a\Upsilon^{5/2}+c_{-1\,1}\Upsilon^{3/2}
+3/(10\a)c_{-1\,1}^{2}\Upsilon^{1/2}+c_{-3\,1}\Upsilon^{-1/2}+\cdots,\\
y&=& 5\a/2\Upsilon^{3/2}+3/2 c_{-1\,1}\Upsilon^{1/2}
+3/(20\a)c_{-1\,1}^{2}\Upsilon^{-1/2}
-1/2c_{-3\,1}\Upsilon^{-3/2}+\cdots,\\
\label{-4/5_B1w}
\nonumber
z&=& 75/(8A)\Upsilon^{-2}-15/(2A\a)c_{-1\,1}\Upsilon^{-3}+
9/(2A\a^{2})c_{-1\,1}^{2}\Upsilon^{-4}+\\
&\,\,\,&+\left(-15/(2A\a)c_{-3\,1}-9/(4A\a^{3})c_{-1\,1}^{3}\right)\Upsilon^{-5}
+\cdots.
\eq
The compatibility conditions at $j=-1$ is satisfied since
\be
(_{-4/5}\mathcal{K}_{1}+\mathcal{I}_{3})\mathbf{c}_{-1}=\left(
  \begin{array}{ccc}
     -3/2      & 1           & 0 \\
     15/4      & -1/2        & 2A\a/5 \\
     75/(4A\a) & -15/(2A\a)  & 1 \\
  \end{array}
\right)c_{-1\,1}\left(
                 \begin{array}{c}
                   1\\
                   3/2  \\
                   -15/(2A\a) \\
                 \end{array}
               \right)=\left(
                         \begin{array}{c}
                           0 \\
                           0 \\
                           0 \\
                         \end{array}
                       \right).
\ee
But for $j=-3$ we find
\bq
\nonumber
&&(_{-4/5}\mathcal{K}_{1}+3\mathcal{I}_{3})\mathbf{c}_{-3}=\\
\nonumber &=&\left(
  \begin{array}{ccc}
    1/2       & 1          & 0 \\
    15/4      & 3/2        & 2A\a/5 \\
    75/(4A\a) & -15/(2A\a) & 3 \\
  \end{array}
\right)\left(
                 \begin{array}{c}
                       c_{-3\,1}\\
                   -1/2c_{-3\,1} \\
                   -15/(2A\a)c_{-3\,1}-9/(4A\a^{3})c_{-1\,1}^{3} \\
                 \end{array}
               \right)\\
               &=&\left(
                         \begin{array}{c}
                           0 \\
                           -9/(10\a^{2})c_{-1\,1}^{3} \\
                           -27/(4A\a^{3})c_{-1\,1}^{3} \\
                         \end{array}
                       \right)=P_{-3}.
\eq An eigenvector corresponding to the eigenvalue $j=-3$ is
$\mathbf{v}^{\top}=(-2A\a/15,A\a/15,1)$, and hence we have \be
\mathbf{v}^{\top}\cdot P_{-3}\neq 0, \ee unless $c_{-1\,1}=0$. In
order to satisfy the compatibility condition at $j=-3$ we set
$c_{-1\,1}=0$. The solution (\ref{-4/5_B1x})-(\ref{-4/5_B1w}) with
$c_{-1\,1}=0$ reads \bq \label{-4/5_B1x2}
x&=& \a\Upsilon^{5/2}+c_{-3\,1}\Upsilon^{-1/2}+\cdots,\\
y&=& 5\a/2\Upsilon^{3/2}-1/2c_{-3\,1}\Upsilon^{-3/2}+\cdots,\\
\label{-4/5_B1w2}
z&=&75/(8A)\Upsilon^{-2}-15/(2A\a)c_{-3\,1}\Upsilon^{-5}+\cdots \eq
and it is a particular solution containing two arbitrary constants.
As $S\equiv 1/\Upsilon\rightarrow\infty$, we conclude that \be
a\rightarrow \infty, \quad a'\rightarrow\infty, \quad
\rho\rightarrow \infty, \ee and we can therefore avoid the
finite-distance singularity in this case.

Next we examine the balance $_{\gamma}\mathcal{B}_{2}$ when
$\gamma>-1/2$. For $\gamma=0$, we have that
$\textrm{spec}(_{0}\mathcal{K}_{2})=\{-1,0,-2\}$, and hence we
substitute \be x=\Sigma_{j=0}^{-\infty}c_{j1}\Upsilon^{j+1},\quad
y=\Sigma_{j=0}^{-\infty}c_{j2}\Upsilon^{j},\quad
z=\Sigma_{j=0}^{-\infty}c_{j3}\Upsilon^{j-2}, \ee and find: \bq
\label{0_B2x}
x&=& \a\Upsilon+c_{-1\,1}-A\a/3c_{-2\,3}\Upsilon^{-1}+\cdots,\\
y&=& \a+A\a/3 c_{-2\,3}\Upsilon^{-2}+\cdots,\\
\label{0_B2w} z&=&  c_{-2\,3}\Upsilon^{-4}+\cdots. \eq The
compatibility conditions at $j=-1$ and $j=-2$ are indeed satisfied
since \be (_{0}\mathcal{K}_{2}+\mathcal{I}_{3})\mathbf{c}_{-1}=
\left(
  \begin{array}{ccc}
     0 & 1  & 0 \\
     0 & 1  & -2A\a/3 \\
     0 & 0  & -1 \\
  \end{array}
\right)c_{-1\,1}\left(
                 \begin{array}{c}
                   1\\
                   0  \\
                   0 \\
                 \end{array}
               \right)=\left(
                         \begin{array}{c}
                           0 \\
                           0 \\
                           0 \\
                         \end{array}
                       \right),
\ee
and
\be
(_{0}\mathcal{K}_{2}+2\mathcal{I}_{3})\mathbf{c}_{-2}= \left(
  \begin{array}{ccc}
    1 & 1  & 0 \\
     0 & 2 & -2A\a/3 \\
     0 & 0  & 0 \\
  \end{array}
\right)c_{-2\,3}\left(
                 \begin{array}{c}
                   -A\a /3\\
                   A\a/3  \\
                   1 \\
                 \end{array}
               \right)=\left(
                         \begin{array}{c}
                           0 \\
                           0 \\
                           0 \\
                         \end{array}
                       \right).
\ee As $S\equiv 1/\Upsilon\rightarrow\infty$, we conclude that \be
a\rightarrow \infty, \quad a'\rightarrow\infty, \quad
\rho\rightarrow \infty, \ee and the finite-distance singularity in
shifted at an infinite distance.

We now move on to the balance $_{-1/2}\mathcal{B}_{3}$, $r>-2$. In this case
this balance has two negative $\mathcal{K}$-exponents. If we choose the value
$r=0$, then the spectrum is found to be
\be
\textrm{spec}(_{-1/2}\mathcal{K}_{3})=\{-1,0,-2\},
\ee
and so inserting the forms
\be
x=\Sigma_{j=0}^{-\infty}c_{j1}\Upsilon^{j+1},\quad
y=\Sigma_{j=0}^{-\infty}c_{j2}\Upsilon^{j},\quad
z=\Sigma_{j=0}^{-\infty}c_{j3}\Upsilon^{j},
\ee
we obtain
\bq
\label{-1/2_B3x}
x&=& \a\Upsilon+c_{-1\,1},\\
y&=& \a,\\
\label{-1/2_B3w} z&=&  c_{-2\,3}\Upsilon^{-2}+\cdots,
\eq
which validates the compatibility conditions at $j=-1$ and $j=-2$ since
\be (_{-1/2}\mathcal{K}_{3}+\mathcal{I}_{3})\mathbf{c}_{-1}= \left(
  \begin{array}{ccc}
      0 & 1  &  0 \\
      0 & 1  &  0 \\
      0 & 0  & -1 \\
  \end{array}
\right)c_{-1\,1}\left(
                 \begin{array}{c}
                   1 \\
                   0 \\
                   0 \\
                 \end{array}
               \right)=\left(
                         \begin{array}{c}
                           0 \\
                           0 \\
                           0 \\
                         \end{array}
                       \right),
\ee
and
\be (_{-1/2}\mathcal{K}_{3}+2\mathcal{I}_{3})\mathbf{c}_{-2}= \left(
  \begin{array}{ccc}
    1 & 1  & 0 \\
     0 & 2  & 0 \\
     0 & 0  & 0 \\
  \end{array}
\right)c_{-2\,3}\left(
                 \begin{array}{c}
                   0 \\
                   0 \\
                   1 \\
                 \end{array}
               \right)=\left(
                         \begin{array}{c}
                           0 \\
                           0 \\
                           0 \\
                         \end{array}
                       \right).
\ee We see that as $S\equiv 1/\Upsilon\rightarrow\infty$, \be
a\rightarrow c_{-1\,1}, \quad a'\rightarrow\a, \quad \rho\rightarrow
\infty, \quad \a\neq 0, \ee so that the balance
$_{-1/2}\mathcal{B}_{3}$ for $r>-2$ also offers the possibility of
escaping the finite-distance singularities. Hence in such cases we
find a regular (singularity free) evolution of the brane as it
travels in the bulk filled with the type of matter considered above.
\section{Conclusions}
We have studied the dynamical `evolution' of a braneworld that consists of a
three-brane embedded in a five-dimensional bulk spacetime filled with a `perfect fluid'
possessing a general equation of state $P=\gamma\rho$,
characterized by the constant parameter $\gamma$.

For a flat brane we find that it is possible to have within finite distance from
the brane the analogous type of collapse singularity met previously in \cite{ack1}.
We call this a collapse type I singularity and describe its nature using
the behavior of the warp factor, its derivative and the
density of the fluid, as $a\rightarrow 0$, $a'\rightarrow \infty$,
$\rho\rightarrow\infty$.
In \cite{ack1}, the bulk matter was modeled by a scalar field and
this singularity was the \emph{only} type possible.
Here we showed that when the scalar field is replaced with a perfect fluid,
in addition to that singularity which appears inevitably in all flat brane solutions with $\gamma>-1/2$,
there are two other new types (for a flat brane):
The first one is the very distinct big rip singularity which
occurs with $a\rightarrow\infty$, $a'\rightarrow -\infty$,
$\rho\rightarrow\infty$ and only when a phantom type equation of state with $\gamma<-1$ is considered.
The second one is a collapse type IIc singularity which may be described by
the behavior $a\rightarrow 0$, $a'\rightarrow \alpha$
and $\rho\rightarrow\infty$.
We note that this latter singularity is less general than the collapse type I and
the big rip singularities and it arises only when $\gamma=-1/2$.
Besides these singular solutions, we found the surprising result of flat branes without finite-distance singularities in the region $-1<\gamma\le -1/2$. Moreover, for $\gamma=-1/2$ the solution has the sudden behavior with $a$ and $a'$ finite and vanishing density $\rho\to 0$~\cite{ba04}.

In contrast to the bulk scalar field case where all curved brane solutions were regular,
now we found also singular such solutions. The possible corresponding
finite-distance singularities are the ones comprising the collapse type II class. These are
singularities with $a\rightarrow 0$, $a'\rightarrow \alpha$
and $\rho\rightarrow 0, \rho_{s},\infty$ (corresponding to types IIa, b and c
respectively).
The interesting feature of this class of singularities is that it
allows the `energy' leak into the extra dimension to vary and be
monitored each time by the $\gamma$ parameter that defines the type
of fluid; they all arise in the region $\gamma\le -1/2$.
On the other hand, we showed that for a curved brane the
possibility of avoiding the finite-distance singularities that was
offered in \cite{ack1} is still valid here, but only in the region $\gamma\ge -1/2$.

For illustration, we present a summary of all different behaviors we found for flat and curved branes in the table below, using the notation for the various singularities introduced in Section 2 after eq.~(\ref{constraint3}) and the balances (\ref{preveq})-(\ref{Bfive}).
\begin{center}
\begin{tabular}{|c|c|c|c|c|}
\hline
equation of state &\mco{2}{c|}{flat brane} & \mco{2}{c|}{curved brane} \\
\hline
$P=\gamma\rho$ & type & balance & type & balance \\ \hline
$\gamma>-1/2$ & singular type I & $_{\gamma}\mathcal{B}_{1}$ & {\em regular} &
$_{\gamma}\mathcal{B}_{2}$ at $\infty$ \\ \hline
$\gamma=-1/2$ & singular IIc & $_{-1/2}\mathcal{B}_{4}$ & {\em regular} &
$_{-1/2}\mathcal{B}_{3}\,$, $r>-2$ at $\infty$ \\
 & {\em regular} sudden & $_{-1/2}\mathcal{B}_{5}$ & singular IIc &
 $_{-1/2}\mathcal{B}_{3\,}$, $r<-2$ or $_{-1/2}\mathcal{B}_{4}$ \\ \hline
$-1<\gamma<-1/2$ & {\em regular} & $_{\gamma}\mathcal{B}_{1}$ at $\infty$ & singular IIa,b,c &
$_{\gamma}\mathcal{B}_{2}$ \\ \hline
$\gamma=-1$ & no solution & & singular IIb & $_{\gamma}\mathcal{B}_{2}$ \\ \hline
$\gamma<-1$ & singular big rip & $_{\gamma}\mathcal{B}_{1}$ & singular IIa &
$_{\gamma}\mathcal{B}_{2}$ \\
\hline
\end{tabular}
\end{center}

An open question is whether there exist physical constraints on
$\gamma$ analog to the weak and strong energy conditions of matter
perfect fluid in ordinary cosmology. A related question is to find
possible field theory realizations of the `exotic' regions of
$\gamma\le -1/2$, where interesting solutions with unexpected
behavior were found. The most important issue of course is to
clarify the possibility of singularity avoidance at finite distance
in flat brane solutions. There is no reason why  the non-singular
behavior for flat branes discovered here  should not persist for
arbitrary values of the brane tension and, indeed, it is to be
expected that only particular asymptotic modes of behavior, that is
specific detailed forms of asymptotic solutions, would depend on
such values. 
Thus, the self-tuning mechanism appears to be a property of a general
(non-singular) flat brane solution, that depends on two arbitrary constants in the region $-1<\gamma<-1/2$
(three for the general solution with sudden behavior when $\gamma=-1/2$).
Similarly, as we have shown here, the existence of a
singularity may be independent of the sign of
the scalar curvature (as long as the latter remains non zero for curved
branes), but the particular way of asymptotic approach to the
singularity is sensitive to that sign and it may therefore change
with different values of brane tension.


It would also be interesting to further investigate whether the properties of finite-distance
singularities (and their possible avoidance) encountered here
continue to emerge in more general systems, such as the case in
which a scalar field \emph{coexists} with a perfect fluid in the bulk~ \cite{ack3}. The analysis
of this more involved case may also shed light to the factors that control how
these two bulk matter components compete on approach to the singularity, or even
predict new types of singularities that might then become feasible,
as well as possible situations where they can be avoided.

\section*{Acknowledgements}
S.C. and I.K. are grateful to CERN-Theory Division, where part of
their work was done, for making their visits there possible and for
allowing them to use its excellent facilities. The work of I.A. was
supported in part by the European Commission under the  ERC Advanced
Grant 226371 and the contract PITN-GA-2009-237920 and in part by the
CNRS grant GRC APIC PICS 3747.


\end{document}